\documentclass{PoS}
\usepackage{caption}
\usepackage{lipsum}
\usepackage{floatrow}
\usepackage[outercaption]{sidecap}    
\title{The broad line region and dust torus structure of the Seyfert 1 WPVS48}

\ShortTitle{Bowl-shaped dust torus for the Seyfert 1 WPVS48}

\author{C. Sobrino Figaredo\\
        Astronomisches Institut - Ruhr Universit\"{a}t Bochum, Germany\\
        E-mail: \email{catso@astro.rub.de}}
\author{Francisco Pozo Nu\~{n}ez\\
        Haifa Research Center for Theoretical Physics and Astrophysics, Israel.\\
        Astronomisches Institut - Ruhr Universit\"{a}t Bochum, Germany\\ 
}
\author{Michael Ramolla\\
        Astronomisches Institut - Ruhr Universit\"{a}t Bochum, Germany
}
\author{Martin Haas\\
        Astronomisches Institut - Ruhr Universit\"{a}t Bochum, Germany
} 
\author{Rolf Chini\\
        Astronomisches Institut - Ruhr Universit\"{a}t Bochum, Germany
}
\author{Klaus-Werner Hodapp\\
        IfA, Hawaii, USA
}
\author{Steve Willner\\
        Harvard-Smithsonian Center for Astrophysics, Cambridge, USA
}
\author{Matt Ashby\\
        Harvard-Smithsonian Center for Astrophysics, Cambridge, USA
}

\abstract{Optical and near-mid-infrared reverberation mapping data obtained at Universit\"{a}tssternwarte Bochum in Chile and with the Spitzer Space Telescope allow us to explore the geometry of both the H$\alpha$ BLR and the dust torus for the nearby Seyfert 1 galaxy WPVS\,48. On average, the H$\alpha$ variations lag the blue AGN continuum by about 18 days, while the dust emission variations lag by 70 days in
the J+K and by 90 days in the L+M bands. The IR echoes are sharp, while the H$\alpha$ echo is smeared. This together favours a bowl shaped toroidal geometry where the dust sublimation radius is defined by a bowl surface, which is virtually aligned with a single iso-delay surface, thus leading to the sharp IR echoes. The BLR clouds, however, are located inside the bowl and spread over a range of iso-delay surfaces, leading to a smeared echo.}

\FullConference{Revisiting narrow-line Seyfert 1 galaxies and their place in the Universe - NLS1 Padova\\
		9-13 April 2018 \\
		Padova Botanical Garden, Italy}

\begin{document}

\section{Introduction}
The paradigm of active galactic nuclei (AGN) consists of an accreting black hole (BH) at the centre of the galaxy and an accretion disk (AD), which is surrounded by the broad line region (BLR) composed of clouds of ionized gas moving rapidly in the BH potential. Clouds located further out move slower and form the narrow line region (NLR). In the equatorial plane, the BLR is surrounded  by a (clumpy) molecular dust torus \cite{Antonucci-1985}. \\

\begin{SCfigure}
\hspace*{-0.5cm}
\includegraphics[width=80mm]{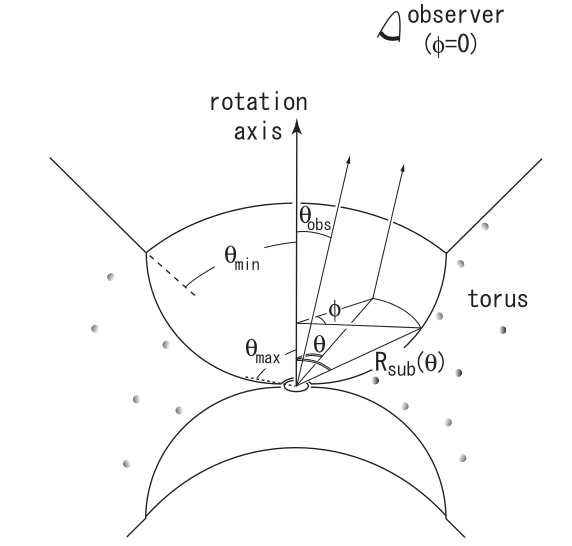}
\caption{Bowl-shaped torus model from \cite{Kagawuchi-2010}. The dust sublimation radius $R_{sub}$ depends on the polar angle $\Theta$. $R_{sub} (\Theta)$ has a convex form resembling a bowl. The bowl dust torus covers a minimum opening angle
$\Theta_{min}$ and a maximum opening angle of $\Theta_{max}$. The AD emission anisotropy allows the torus's inner region to closely approach the central BH and AD in the equatorial plane.}
\label{fig:bowl_model}
\end{SCfigure}

The dust sublimation radius $R_{sub}$ is defined as the smallest distance from the centre of the galaxy to the dust torus' surface \cite{Barvanais-1982}. However, \cite{Suganuma-2006} analysed data for a sample of Seyfert galaxies and found that the dust torus radius $R_{\tau_K}$ is almost 3 times smaller than the expected $R_{sub}$. This systematic deviation can be explained by a bowl-shaped dust torus geometry, proposed by \cite{Kagawuchi-2010}. In this model, the AD emission is assumed to be non-isotropic, declining from the polar direction towards the equatorial plane, so that the sublimation radius defines a convex bowl-shaped rim, which smoothly transitions into the AD (see Figure \ref{fig:bowl_model}). Assuming the model in \cite{Goad-2012}, the BLR clouds cover the bowl-shaped dust rim and they occupy the region between AD and dust torus.\\

Here, we explore the central geometry of the nearby Seyfert 1 galaxy WPVS\,48 \cite{vernon-cetty} and discuss a bowl-shaped BLR/dust torus model configuration. Previous data has already been analysed in \cite{pozo-nunez-2014}, where WPVS\,48 was shown to exhibit strong optical and near-infrared variations. It is located at redshift $z = 0.037$ and the FWHM of H$\alpha$ is 937 km/s. Since the central region of AGN cannot be spatially resolved on images, the reverberation mapping (RM) technique is applied.

\section{AD, BLR and Dust Light curves}

WPVS\,48 has been monitored for 6 months using the VYSOS-6, BEST-II, BMT and IRIS telescopes located at the Bochum observatory near Cerro Armazones, the future location of the ESO
Extreme Large Telescope (ELT) in Chile\footnote{http://www.astro.ruhr-uni-bochum.de/astro/oca/}. The broad band optical filters $B$, $V$ and $R$ are employed to analyse the AGN continuum. A narrow band filter centered in 680\,nm ($NB680$) monitors the broad $H\alpha$ emission line variation and the NIR filters ($J$, $K$) the variations of the hot dust. Additional MIR data obtained with the Spitzer Space telescope ($L = 3.6\,\mu m$ and $M = 4.5\,\mu m$) is used to investigate cooler dust.
\begin{figure}
\hspace*{-0.5cm}
\includegraphics[width=145mm]{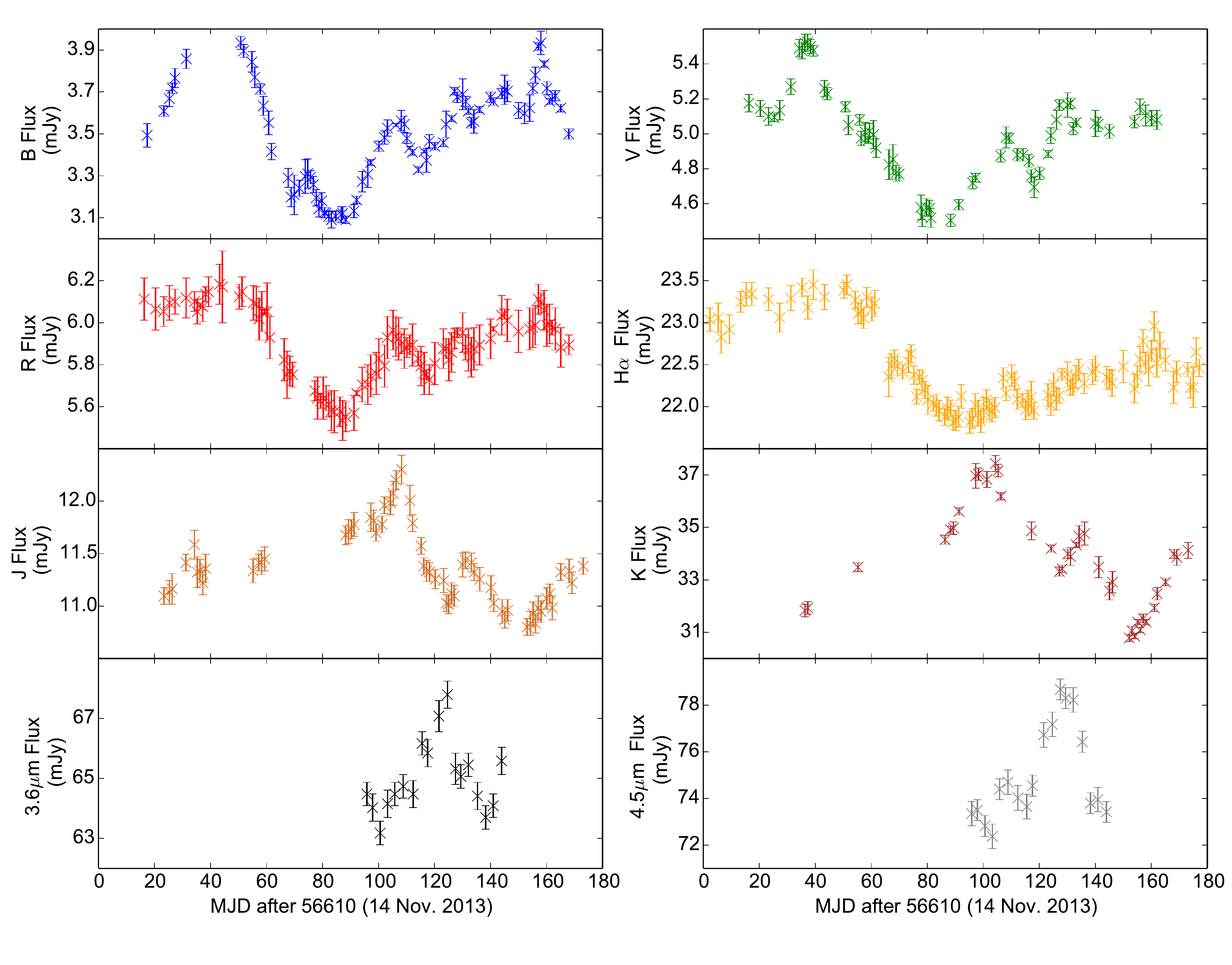}
\caption{WPVS48 calibrated light curves, all corrected for galactic foreground extinction.}
\label{fig:lc}
\end{figure}
The flux calibrated light curves  are shown in Figure \ref{fig:lc}.

The $B$-band light curve shows a strong flux increase until MJD $\sim$ 56610 + 40, then the flux decreases by almost 30\% and reaches a minimum at MJD $\sim$ 56610 + 85; afterwards, it begins to increase again. The $V$ and $R$-band light curves show a similar variability pattern but the amplitude variation in $R$ is less pronounced, probably due to a stronger host galaxy contribution. Since $NB680$ contains the contribution of the red continuum, a synthetic $H\alpha$ light curve is created by calculating $H\alpha = NB680 - R$ \cite{Haas-2011}, which varies by less than 10\%. The $H\alpha$ light curve shows a decline shifted by $\sim$ 15 days from $B$ and $V$. After the minimum is reached, the increase is less pronounced suggesting that the transfer function contains long lag contributions. The $J$ and $K$ light curves again show the same variability pattern as do $B$ and $V$ (shifted by $\sim$ 70 days) and a large amplitude (almost 30\%). The $L$ and $M$-bands also show pronounced sharp variations shifted by \mbox{$\sim$ 20} days against $J$ and $K$.\\

\begin{figure}
\includegraphics[width=105mm]{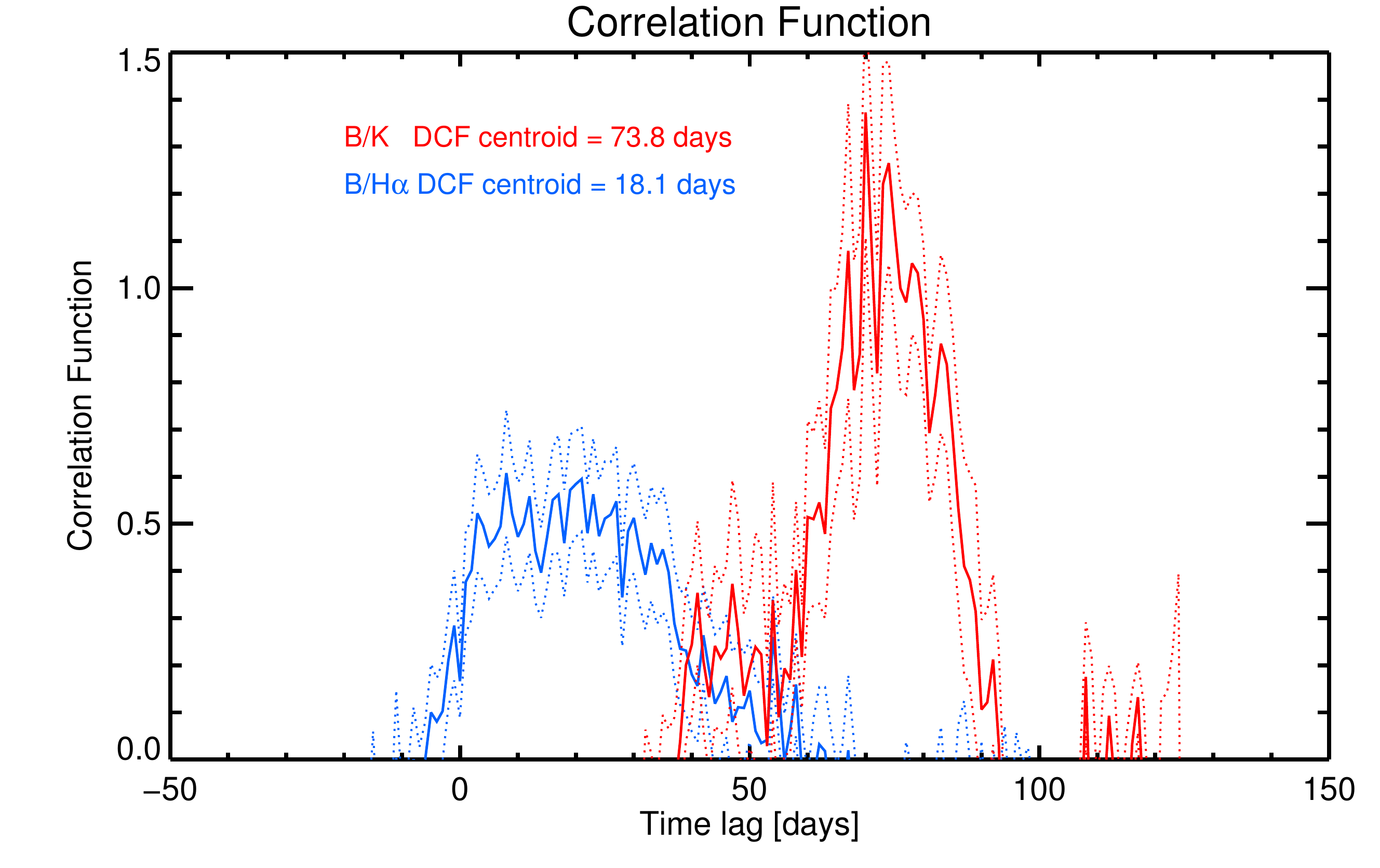}
\caption{Correlation function for $B/H\alpha$ and $B/K$ plotted with blue and red solid lines, respectively. The dashed lines indicate the error range
($\pm$ 1$\sigma$) around the cross correlation.}
\label{fig:dcf}
\end{figure}

In order to obtain the time delay between the AGN-continuum and BLR/dust light curves we compute the Discrete Correlation Function (DCF) \cite{Edelson-Krolik}. Figure \ref{fig:dcf} shows the DCF function for $B/H\alpha$ and $B/K$. We find an average time lag between $B/H\alpha$ of 18 days with a broad correlation function (between 0 and 40 days and a tail up to 60 days). The NIR filters show an average time lag of 70 days ($B/J-K$) and 90 days for MIR filters ($B/L-M$). The NIR light curves show a narrower and a larger correlation value than the $H\alpha$ light curve (see example in Figure \ref{fig:dcf}), which is in accordance with a larger amplitude variation in the NIR filters than in the $H\alpha$.

\section{Discussion: Bowl-shaped dust model}

The observation of a smeared $H\alpha$ echo light curve (broad DCF) and sharp NIR-MIR echo light curves (narrower DCF) can be explained by a bowl-shaped central AGN geometry.\\

The fact that the hot and cooler dust light curves are quite sharp indicates that the observer looks at the system nearly face-on. In addition, the explanation for such a sharp echo could be that the emitting rim of a bowl-shaped dust torus is almost aligned with an iso-delay contour. If the $H\alpha$ BLR clouds are located above the bowl-shaped dust rim, they will be spread across a range of iso-delay contours and therefore will produce a smeared echo. This possible geometry for the central region of WPVS48 will be discussed in a forthcoming publication (Sobrino Figaredo et al., in preparation).\\

\section*{Acknowledgements}

This conference has been organized with the support of the
Department of Physics and Astronomy ``Galileo Galilei'', the 
University of Padova, the National Institute of Astrophysics 
INAF, the Padova Planetarium, and the RadioNet consortium. 
RadioNet has received funding from the European Union's
Horizon 2020 research and innovation programme under 
grant agreement No~730562.

\end{document}